\documentclass[aps,prb,reprint,groupedaddress,superscriptaddress,longbibliography]{revtex4-2}
\usepackage{amsmath}
\usepackage{amssymb}
\usepackage{graphicx}
\usepackage{mathrsfs}
\usepackage{subfigure}
\usepackage{comment}
\usepackage{ntheorem}
\usepackage{tabularx}
\usepackage{xcolor}
\usepackage{bbm}
\usepackage{dsfont}
\usepackage{bbold}
\usepackage[hidelinks]{hyperref}
\newcommand{\wsx}{\color {black}}
\newcommand{\zb}{\color {black}}
\newcommand{\zbb}{\color {black}}

\begin{document}
\newtheorem{Definition}{Definition}[subsection]
   \title{Theory for the spectral splitting exponent of exceptional points}
   \author{Shu-Xuan Wang}
   \email{wangshx65@mail.sysu.edu.cn}
   \affiliation{Guangdong Provincial Key Laboratory of Magnetoelectric Physics and Devices, State Key Laboratory of Optoelectronic Materials and Technologies, School of Physics, Sun Yat-sen University, Guangzhou 510275, China}
   \author{Zhongbo Yan}
   \email{yanzhb5@mail.sysu.edu.cn}
   \affiliation{Guangdong Provincial Key Laboratory of Magnetoelectric Physics and Devices, State Key Laboratory of Optoelectronic Materials and Technologies, School of Physics, Sun Yat-sen University, Guangzhou 510275, China}

   \date{\today}

   \begin{abstract}
      Exceptional points (EPs), singularities in non-Hermitian systems where eigenvalues and eigenstates coalesce, 
      exhibit a dramatically enhanced response to perturbations compared to Hermitian degeneracies. 
      This makes them exceptional candidates for sensing applications. The spectral splitting of 
      an $N$th-order EP scales with perturbation strength $\epsilon$ over a wide range, from $\epsilon$ to $\epsilon^{1/N}$. 
      Although the exact scaling exponent can be determined in principle by solving the characteristic equation, this approach becomes analytically intractable for large $N$ and often fails to yield useful physical insight. 
      In this work, we develop a theory to directly predict the scaling 
      exponent from the matrix positions of the perturbation. By using the Jordan block structure of 
      the unperturbed Hamiltonian, we show that the splitting exponent can be analytically determined when the 
      matrix positions of the perturbation satisfy some specific conditions.
      Our analytical framework provides a useful design principle for engineering 
      perturbations to achieve a desired spectral response, facilitating the development of 
      EP-based sensors. 
   \end{abstract}

   \maketitle

   \section{Introduction}
        Recently, non-Hermitian {\zb Hamiltonians} have garnered considerable attention~\cite{RevModPhys.93.015005, ashida2020non} as an effective framework for describing open quantum {\zb and classical} systems. In such systems, the effective Hamiltonian becomes non-Hermitian due to gain and loss~\cite{Rotter_2009, PhysRevLett.70.2273}. Notably, the introduction of non-Hermiticity gives rise to {\zb numerous phenomena without} Hermitian counterparts, such as the emergence of exceptional points (EPs)~\cite{W.D.Heiss_2004,berry2004physics}, the  
        non-Bloch band theory~\cite{PhysRevLett.121.086803, PhysRevLett.123.066404, PhysRevB.101.195147, PhysRevX.14.021011, HU202551, zhang2022universal, PhysRevLett.133.216401, PhysRevB.107.115412}, non-Hermitian skin effects~\cite{PhysRevLett.121.086803, PhysRevLett.123.066404, PhysRevLett.124.056802, PhysRevLett.125.126402, PhysRevLett.124.086801, PhysRevLett.125.226402, zhang2022universal, PhysRevLett.125.186802, PhysRevB.107.115412, PhysRevB.109.L081108, PhysRevB.108.L060204, PhysRevB.106.L241112, li2021impurity, PhysRevResearch.5.033058, PhysRevB.108.075122, PhysRevB.108.205423, PhysRevLett.127.256402, PhysRevLett.131.116601, PhysRevLett.127.116801, PhysRevB.111.L140201, li2020critical, PhysRevB.104.165117, PhysRevB.107.155430, li2024observation, PhysRevLett.123.016805, PhysRevB.102.241202, PhysRevB.103.045420,Chen_2024,manna2023inner}, novel non-Hermitian topological phases~\cite{PhysRevX.8.031079, PhysRevX.9.041015, PhysRevLett.120.146402, PhysRevB.110.L121401, PhysRevB.99.201103, PhysRevLett.116.133903, PhysRevLett.121.136802, PhysRevLett.126.216405, PhysRevLett.122.076801, PhysRevB.102.205118, PhysRevLett.123.016805, PhysRevB.99.081302, PhysRevB.102.241202, PhysRevB.103.045420,PhysRevB.111.085148,PhysRevB.111.094109,PhysRevB.111.195419}, edge bursts~\cite{PhysRevLett.128.120401, PhysRevB.108.235422, PhysRevLett.133.070801}, non-Bloch $\mathcal{PT}$ symmetry breaking~\cite{PhysRevLett.132.050402, PhysRevLett.126.230402}, and  {\zb n}on-Hermitian invisibility~\cite{PhysRevA.82.032111, PhysRevB.106.094205}.

        EPs correspond to degeneracies in non-Hermitian energy spectra~\cite{Heiss_2012, ding2022non, PhysRevResearch.2.043045, PhysRevLett.126.086401, PhysRevB.105.075420, PhysRevB.104.L201104, PhysRevB.104.L161116, PhysRevB.104.L161117, PhysRevLett.123.066405, gohsrich2024, PhysRevLett.128.010402, denner2021exceptional, PhysRevB.110.L201104, wang2025, PhysRevResearch.6.043023, PhysRevLett.127.186602, PhysRevLett.118.045701, PhysRevB.99.081102, PhysRevA.98.042114, PhysRevB.99.161115, PhysRevLett.126.010401,PhysRevB.110.L020104,PhysRevLett.134.133801,18gg-gvzc,PhysRevResearch.4.023130,bid2025fragmentedexceptionalpointsbulk,PhysRevResearch.7.023062}, {\zbb and share intriguing connections with Newton polygons and tropical geometry~\cite{jaiswal2023characterizing,banerjee2023tropical}}.
        These spectral singularities give rise to a plethora of exotic physical phenomena, such as unidirectional invisibility~\cite{Lin2011unidirectional,Peng2014}, 
        topological energy transfer~\cite{Doppler2016,Xu2016} and laser mode selectivity~\cite{Hossein2014,Feng2014}.  Due to the rich physics they enable, EPs have been realized experimentally across numerous platforms, including photonic crystals~\cite{ozdemir2019parity, regensburger2012parity, zhou2023observation, kim2016direct}, exciton-polariton systems~\cite{gao2015observation}, cavities~\cite{PhysRevX.6.021007, PhysRevLett.103.134101}, metamaterials~\cite{Park2020} and synthetic systems~\cite{yang2023realization}. 
        
        A salient property of EPs is their sensitive spectral response to perturbations, which is 
        rooted in their singular wave-function structure. 
        Unlike Hermitian degeneracies, where 
        eigenstates remain orthogonal, EPs are characterized by the simultaneous 
        coalescence of both eigenvalues and their corresponding eigenstates.
        This fundamental distinction leads to significantly different responses 
        to perturbations. Specifically, under a perturbation of strength $\epsilon$, 
        the spectral splitting for a Hermitian degeneracy scales at most linearly with 
        $\epsilon$ ($\Delta E \sim \epsilon$). In stark contrast, the splitting for an $N$th-order EP
        (where $N$ eigenstates coalesce)
        can scale as $\Delta E \sim \epsilon^{1/N}$~\cite{PhysRevA.101.033820, chen2017exceptional, hodaei2017enhanced}. The ratio of these responses, $\epsilon^{1/N}/\epsilon = \epsilon^{(1-N)/N}$, 
        diverges in the limit $\epsilon \rightarrow 0$. This indicates that the response of an EP to a perturbation 
        can be much stronger than that of any Hermitian degeneracy, a property that makes EPs 
        exceptional candidates for enhancing sensitivity in sensing applications~\cite{PhysRevLett.112.203901, PhysRevA.93.033809, chen2017exceptional, hodaei2017enhanced, PhysRevLett.123.213901, PhysRevLett.122.153902, PhysRevApplied.12.024002, PhysRevLett.125.240506, PhysRevA.101.053846, PhysRevResearch.6.033284}. {\zbb The ultimate performance in practical devices, however, is limited by noise, which bounds the signal-to-noise ratio~\cite{lau2018fundamental,Chen_2019}. Strategies to overcome this limitation remain an active area of research~\cite{Kononchuk2022}.
        A deeper understanding of the spectral splitting behavior can shed new light on these key problems. }
        
        While the strongest response, $\Delta E \sim \epsilon^{1/N}$, should in principle yield optimal sensor performance, 
        this scaling is not guaranteed for an arbitrary perturbation~\cite{PhysRevLett.127.186601, PhysRevLett.134.153601, wu2021high,PhysRevB.99.241403}. 
        For an $N$th-order EP, the exponent $\alpha$ characterizing 
        the leading-order splitting (defined by $\Delta E \sim \epsilon^{\alpha}$) can range from 
        $\alpha = 1$ to $\alpha = 1/N$. Specially,  the exponent takes the general form $\alpha = 1/k$, 
        where $k$ is an integer satisfying $1 \leq k \leq N$. The integer $k$ corresponds to the number 
        of eigenstates that escape from the coalescence. Although the exponent $\alpha$ can 
        always be determined in principle by solving the characteristic equation, 
        this approach becomes analytically intractable for large $N$. Furthermore, numerical solutions, 
        while feasible, often fail to provide physical insight into the relationship between the perturbation's 
        form and the resulting exponent. Therefore, an analytical theory capable of predicting $\alpha$ without 
        solving for the full spectrum would be highly valuable, as it would enable the design of perturbations 
        that yield a desired splitting scaling.

        In a previous work, we explored the spectral splitting of infernal points---a class of EP whose order 
        scales with system size---under perturbations~\cite{wang2025}. Because the Hamiltonian at these points takes 
        a Jordan form, we were able to analytically derive the condition for achieving 
        the strongest possible response ($\alpha=1/N$). Intriguingly, the condition exhibits a simple relationship 
        to the perturbation term's positions in the matrix representation. Given that the Jordan block 
        structure is a universal feature of Hamiltonians at EPs, we here extend this theory to general EPs 
        of arbitrary order. We demonstrate that the splitting exponent $\alpha$ for a generic response 
        can be determined analytically when the perturbation matrix satisfies specific algebraic conditions. 
        This establishes a direct connection between the perturbation's matrix structure and 
        the exponent $\alpha$, with our previous result emerging as a special case~\cite{wang2025}.

        The rest of this paper is organized as follows. In Sec.~\ref{II}, we introduce 
        our general theoretical framework.  In Sec.~\ref{III}, 
        we analyze four distinct classes of perturbations and derive their characteristic 
        spectral splitting behavior. In Sec.~\ref{IV}, we demonstrate the validity of 
        our general framework using a fourth-order EP as a concrete example. Finally, 
        we present a discussion and our conclusions in Sec.~\ref{V}.

    \section{General framework}\label{II}
    
       We consider a general $N\times N$ non-Hermitian Hamiltonian $H_{\rm N}$ which supports an
       $N$th-order EP for appropriate parameters. Without loss of generality, we assume 
       the EP occurs at energy $E_0$. At this degeneracy, $H_{\rm N}$ possesses a single eigenstate, 
       $|u_0\rangle$, satisfying $H_{\rm N}|u_0\rangle=E_{0}|u_0\rangle$. 
       The complete $N$-dimensional Hilbert space is spanned by the Jordan chain, comprising this eigenstate and its $N-1$ associated vectors: ${|u_0\rangle, |u_1\rangle, \cdots, |u_{N-1}\rangle}$.
       Correspondingly, the system has a single left eigenstate, $|v_0\rangle$, 
       satisfying $H_{\rm N}^{\dagger}|v_0\rangle = E_{0}^{*}|v_0\rangle$, along 
       with $N-1$ left associated vectors, $ |v_1\rangle, \cdots, |v_{N-1}\rangle$. 
       Owing to the Hamiltonian's defectiveness, the standard biorthogonality 
       condition $\langle v_{i} | u_{j} \rangle = \delta_{ij}$ breaks down. 
       Instead, the vectors obey the Jordan chain structure~\cite{seyranian2003multiparameter}, 
       characterized by the relation $\langle v_{i} | u_{j} \rangle = \delta_{i+j, N-1}$ 
       (A brief review is provided in Appendix A).

        We introduce a perturbation to the system, $H_p = \epsilon H_1$, where $\epsilon$ 
        is a dimensionless parameter that characterizes the perturbation strength. Under 
        this perturbation, we assume that the original eigenenergy $E_0$ shifts to $E_0 + \lambda$, and 
        the modified eigenstate $|u\rangle$ can be expanded as
           \begin{equation}
              | u \rangle =  | u_0 \rangle + \sum_{i=1}^{N-1} s_i | u_i \rangle,
              \label{1}
           \end{equation}
        where $s_i$ denotes coefficients. Accordingly,  the Schr\"{o}dinger equation for the perturbed system is
           \begin{equation}
              \left[ H_{\rm N} + H_p - E_{0} - \lambda \right] | u \rangle = 0.
              \label{2}
           \end{equation}
         Substituting Eq.~\eqref{1} into Eq.~\eqref{2} and applying the Jordan chain structure from Eq. (A1), we obtain
           \begin{equation}
              \sum_{i=1}^{N-1} s_i | u_{i-1} \rangle + \left( \epsilon H_1 - \lambda \right) \left[| u_0 \rangle + \sum_{i=1}^{N-1} s_i | u_i \rangle \right]  = 0.
              \label{3}
           \end{equation}
         Left-multiplying Eq.~\eqref{3} successively by $\langle v_0 |$, $\langle v_1 |$, $\cdots$, $\langle v_{L-1} |$ 
         and applying the orthogonal normalization condition given in Eq.~\eqref{A3}, we obtain
           \begin{equation}
              \begin{split}
                 \epsilon \left[ H_1^{0,0} +  \sum_{i=1}^{N-1} s_i H_1^{0,i} \right]  - \lambda s_{N-1} &= 0,
                 \\
                 s_{N-1} + \epsilon \left[ H_1^{1,0} +  \sum_{i=1}^{N-1} s_i H_1^{1,i} \right] - \lambda s_{N-2} &= 0,
                 \\
                 s_{N-2} + \epsilon \left[ H_1^{2,0} +  \sum_{i=1}^{N-1} s_i H_1^{2,i} \right] - \lambda s_{N-3} &= 0,
                 \\
                 \vdots  \qquad   \qquad  \qquad \ \quad &
                 \\
                 s_{2} + \epsilon \left[ H_1^{N-2,0} +  \sum_{i=1}^{N-1} s_i H_1^{N-2,i} \right] - \lambda s_{1} &= 0,
                 \\
                 s_{1} + \epsilon \left[ H_1^{N-1,0} +  \sum_{i=1}^{N-1} s_i H_1^{N-1,i} \right] - \lambda \  &= 0,
              \end{split}
              \label{4}
           \end{equation}
         where $H_1^{i,j}$ represent the element $\langle v_i | H_1 | u_j \rangle$. These equations
         can be written in compact form as:
           \begin{equation}
            \left( \mathcal{J}_{\rm N} + \epsilon \mathcal{H}_1 \right) S = \lambda S,
            \label{5}
           \end{equation}
         where
           \begin{equation}
              \mathcal{J}_{\rm N} = 
                 \begin{pmatrix}
                  0   &  1  &   &   &   \\
                  &  0  & 1 &   &   \\
                  &     & \ddots &  \ddots & \\
                  &     &        &   0   &  1 \\
                  &     &        &        &  0
                 \end{pmatrix}_{N \times N}
                 \label{6}
           \end{equation}
         is the $N \times N $ Jordan block, 
           \begin{equation}
              \mathcal{H}_1 = 
                \begin{pmatrix}
                  H_1^{N-1,0} & H_1^{N-1,1} & \cdots & H_1^{N-1,N-1}  \\
                  H_1^{N-2,0} & H_1^{N-2,1} & \cdots & H_1^{N-2,N-1}  \\
                     \vdots   &    \vdots   & \ddots &    \vdots      \\
                  H_1^{0,0}   &  H_1^{0,1}  & \cdots &  H_1^{0,N-1}
                \end{pmatrix}_{N \times N}
                \label{7}
           \end{equation}
         and $S = \left(1, s_1, s_2, \cdots , s_{N-1} \right)^T$. Hence, 
         solving the characteristic equation given in Eq.~\eqref{2} is equivalent to 
         solving the eigenvalues and eigenvectors of matrix $\mathcal{J} + \epsilon \mathcal{H}_1$.

    \section{Analytical determination of the splitting exponent}\label{III}
    
       Although our approach may appear to merely recast the characteristic equation into a different form, 
       we will demonstrate its utility in determining the splitting exponent. This is possible when 
       the perturbation matrix structure satisfies specific conditions, leveraging the unique 
       properties of the Jordan block structure.

      \subsection{Case 1}
        
        The first case we consider corresponds to the situation where all matrix 
        elements $H_1^{p,q}$ vanish for $ p + q < N$ in $\mathcal{H}_1$. For this case, 
        the general form of the perturbation matrix $\mathcal{H}_1$ reduces to
          \begin{equation}
            \mathcal{H}_1 =
              \begin{pmatrix}
               0 & H_1^{N-1,1} & H_1^{N-1,2} & \cdots & H_1^{N-1,N-1}  \\
               0 & 0 & H_2^{N-1,2} & \cdots & H_1^{N-2,N-1}  \\
                  \vdots   &    \vdots   & \ddots  & \ddots &    \vdots      \\
              0   &  0  & \cdots &  \ddots &  H_1^{1,N-1}   \\
              0  &  0  &  \cdots & \cdots  & 0
              \end{pmatrix}_{N \times N} .
              \label{8}
          \end{equation}
        The matrix $\mathcal{J}_{\rm N} + \epsilon \mathcal{H}_1 - \lambda \mathbbm{1}$ is evidently an upper triangular matrix 
        with all diagonal elements equal to $-\lambda$. Consequently, its characteristic polynomial is simply
          \begin{equation}
            p \left(\lambda, \epsilon,N \right) = \det \left[ \mathcal{J}_{\rm N} + \epsilon \mathcal{H}_1 - \lambda \mathbbm{1} \right] = \left( - \lambda \right)^N.
            \label{9}
          \end{equation}
          The unique root of $p(\lambda, \epsilon, N)$ is $\lambda = 0$, indicating that the perturbation does not shift or split 
          the EP. While this result may seem trivial, it leads to an important conclusion: 
          not every perturbation can influence an EP.

      \subsection{Case 2}
      
         The second case we consider corresponds to the situation where only a single matrix element 
         $H_1^{p,q}$ is non-vanishing.  Based on the conclusion from Case 1, perturbations 
         with $p + q \geqslant N$ leave the EP intact. For $p + q < N$, the eigenenergy 
         shift $\lambda$ satisfies (demonstration is provided in Appendix B)
           \begin{equation}
            \lambda^{N-p} - \epsilon H_1^{p,q} \lambda^q = 0.
            \label{10}
           \end{equation} 
         The nonzero solutions are given by
           \begin{equation}
            \lambda_m = \epsilon^{\frac{1}{N-p-q}} \left( H_1^{p,q} \right)^{\frac{1}{N-p-q}} e^{\frac{2i m \pi}{N-p-q}}  ,
            \label{11}
           \end{equation}
         where $m=1,2, \cdots, N-p-q$.  This result indicates that the spectral splitting 
         scales as $\Delta E \sim \epsilon^{1/(N-p-q)}$. Given that $0 \leqslant p+q < N$, 
         a single-element perturbation can produce any splitting exponent of the form $\alpha = 1/k$ 
         for integers $k$ satisfying $1 \leqslant k \leqslant N$. In particular, the 
         choice $p = q = 0$ yields the strongest possible scaling, $\Delta E \sim \epsilon^{1/N}$, 
         which recovers the result from our previous work~\cite{wang2025}.

      \subsection{Case 3}
      
         The third case we consider corresponds to diagonal perturbations,
         meaning that the only nonzero elements of $\mathcal{H}_1$ are those 
         whose indices satisfy $p + q = N-1$. Accordingly, 
         the matrix $\mathcal{J} + \epsilon \mathcal{H}_1$ takes the form
            \begin{equation}
               \mathcal{J}_{\rm N} + \epsilon \mathcal{H}_1  =
               \begin{pmatrix}
                 \epsilon H_1^{N-1,0}  & \ 1 \ &  &  &   \\
                  & \epsilon H_1^{N-2,1}  & \  1 \ &  &   \\
                       &       &  \ddots  & \ \ddots \ &         \\
                   &   &  &  \ddots & \ 1   \\
                  &    &   &   & \epsilon H_1^{0,N-1} 
               \end{pmatrix}_{N \times N} .
               \label{12}
            \end{equation}
         Apparently, the characteristic polynomial of this case is simply
            \begin{equation}
               p \left(\lambda, \epsilon,N \right) = \prod_{i=m}^{N} \left( \epsilon H_1^{N-m,m-1} - \lambda \right).
               \label{13}
            \end{equation}
          The solutions are
          \begin{equation}
          \lambda_m = \epsilon H_1^{N-m,m-1} \qquad (m=1,2, \cdots, N).
          \label{14}
          \end{equation}
          This result shows that the perturbation induces a linear shift in the eigenvalues. 
          The spectral consequence depends on the matrix elements: identical elements $H_1^{N-m,m-1}$ 
          produce a uniform energy shift, preserving the EP, while non-identical elements lift 
          the degeneracy, reducing the EP's order or completely splitting the EP.

      \subsection{Case 4}
      
         The last case we consider corresponds to the situation where the only nonzero elements lie on the line $p + q = j$.
         For $j \geqslant N$, the EP remains intact, as established in Case 1. For $j = N-1$, the conclusion 
         from Case 3 indicates a linear splitting exponent ($\alpha=1$).

         For $(N-1)/2 < j < N-1$, we find that it is difficult to solve the eigenenergies of the perturbed Hamiltonian. 
         However, the case $j \leqslant (N-1)/2$ is analytically solvable. Under this condition, 
         the spectral modifications are given by (demonstration is provided in Appendix C)
            \begin{equation}
               \lambda_m = \epsilon^{\frac{1}{N-j}} \left( \sum_{k=0}^{j} H_1^{j-k,k} \right)^{\frac{1}{N-j}} e^{\frac{2 i m \pi}{N-j}} ,
               \label{15}
            \end{equation}
         where $m=1,2, \cdots, N-j$. This agrees with the results of Case 2. An important conclusion from Eq.~(\ref{15}) 
         is that all matrix elements $H_1^{p,q}$ with $p+q = j \leqslant (N-1)/2$ contribute to the 
         spectral splitting at the same order. 
         
         The consistent result $\alpha = 1/(N - p - q)$ across the last three cases leads us to conjecture 
         that the splitting exponent is a function only of the index sum $p + q$, {\wsx i.e., $j$}. 
         {\zbb When the perturbation matrix contains terms with different values of $j$, the minimum value of $j$ governs the splitting exponent
         in the extremely weak-perturbation regime.}  In other words, 
         $\alpha$ is governed exclusively by the locations of the nonzero elements
         of the perturbation matrix. {\wsx Numerical verification of this conjecture is provided in Appendix D for the model discussed in Section $\mathbf{IV}$.}

    \section{Application in a concrete model}\label{IV}
    
         While many models have been proposed to realize EPs, we select a 
         non-Hermitian supersymmetric (SUSY) array model~\cite{PhysRevA.101.033820} 
         to illustrate our theoretical framework. This model is ideally suited for 
         this purpose, as it allows for the construction of EPs of any order.
         Specially, the Hamiltonian for an $N$-site non-Hermitian SUSY array is  
           \begin{equation}
                \begin{split}
                  \hat{H}_{\rm L} = &\sum_{m=1}^{N} \left[ \omega_0 +i \gamma \left(N + 1 - 2m\right) \right] c_m^{\dagger} c_m 
                  \\
                  & + \sum_{m=1}^{N-1} J \sqrt{m(N-m)} \left(c_m^{\dagger} c_{m+1} + {\rm h.c.} \right).
                \end{split} 
                \label{16}
           \end{equation}
         This Hamiltonian has a physical interpretation as describing a spin-$(N-1)/2$ particle 
         under the influence of an effective magnetic field given by $\mathbf{B} = (J, 0, i\gamma)$. 
         The characteristic energy $\omega_{0}$ corresponds to the on-resonator frequency of each resonator or
         waveguide, depending on the concrete physical realization of this model. 
    
         The eigenenergies of this Hamiltonian is simply 
         \begin{eqnarray}
         E_{L,n} = \omega_0 + n \sqrt{J^2 - \gamma^2}, 
         \end{eqnarray}
         where $n=-(N-1),-(N-3), \cdots,(N-3), (N-1)$. When $J=\gamma$, 
         all eigenenergies coalesce at $\omega_0$, forming an $N$th-order EP.  
         
         For concreteness, we focus on the case $N=4$, where the Hamiltonian hosts 
         a fourth-order EP at $J=\gamma$. To apply our framework, we first express 
         the Hamiltonian in matrix form. Using the basis $\psi=(c_{1},c_{2},c_{3},c_{4})^{T}$, 
         the Hamiltonian at the EP is $\hat{H}_{4}=\psi^{\dag} H_{4}\psi$, where 
         \begin{equation}
               H_4 = 
                 \begin{pmatrix}
                    \omega_0 + 3 i J  &  \sqrt{3} J  & 0 &  0  \\
                    \sqrt{3} J &  \omega_0 +  i J  & 2 J &  0  \\
                    0  &  2 J  & \omega_0 -  i J  & \sqrt{3} J \\
                    0  &  0  &  \sqrt{3} J & \omega_0 - 3 i J
                 \end{pmatrix}.
                 \label{D1}
            \end{equation} 
         The Hamiltonian can be transformed into its Jordan form via a similarity transformation:
         \begin{equation}
               \mathcal{J}_4 = 
                   \begin{pmatrix}
                     \omega_0 & 1 & 0 & 0  \\
                     0 & \omega_0 & 1 & 0 \\
                     0 & 0 & \omega_0 & 1 \\
                     0 & 0 & 0 & \omega_0
                   \end{pmatrix},
                   \label{D2}
           \end{equation}
         where $\mathcal{J}_4 = U_4^{-1} H_4 U_4$.  
         The transformation matrix $U_4$ and its inverse $U_4^{-1}$ are of the form:
           \begin{gather}
              U_4 = 
                \begin{pmatrix}
                   -i & -\frac{1}{J} & \frac{i}{2 J^2} & \frac{1}{6 J^3}  \\
                   -\sqrt{3} & \frac{2 i}{\sqrt{3} J} & \frac{1}{2 \sqrt{3} J^2} & 0 \\
                   i \sqrt{3} & \frac{1}{\sqrt{3} J}  & 0 & 0 \\
                   1 & 0 & 0 & 0 
                \end{pmatrix},
                \label{D4}
                \\
              U_4^{-1} =  
                \begin{pmatrix}
                  0 & 0 & 0 & 1  \\
                  0 & 0 & \sqrt{3} J & -3 i J \\
                  0 & 2 \sqrt{3} J^2 & -4 i \sqrt{3} J^2 & -6 J^2 \\
                  6 J^3 & -6 i \sqrt{3} J^3 & -6 \sqrt{3} J^3 & 6 i J^3
                \end{pmatrix}.
                \label{D5}
           \end{gather}
         These two matrices can be expressed as 
           \begin{equation}
              \begin{split}
                 U_4 &= \left( | u_0 \rangle, | u_1 \rangle, | u_2 \rangle ,| u_3 \rangle\right),
                 \\
                 U_4^{-1} &= \left( | v_3 \rangle, | v_2 \rangle, | v_1 \rangle ,| v_0 \rangle \right)^{\dagger},
              \end{split}
              \label{D6}
           \end{equation} 
         where $| u_i \rangle$ and $| v_i \rangle$ are the right and left eigenvectors 
         (including associated vectors) of $H_4$, which together satisfy the orthogonal normalization condition 
         in Eq.~\eqref{A3}. 
         
         We now consider five distinct perturbations whose expressions display different levels of complexity. 
         Their explicit expressions are given by
           \begin{equation}
              \begin{split}
                 \hat{H}_{p1} &= \epsilon \frac{J}{6} c_1^{\dagger} c_4 ,
                 \\
                 \hat{H}_{p2} &= \epsilon \frac{J}{4 \sqrt{3}}\left( c_1^{\dagger} c_3  + c_2^{\dagger} c_4\right) ,
                 \\
                 \hat{H}_{p3} &= \epsilon J \left[ \frac{3}{2} c_1^{\dagger} c_4 + \frac{1}{2} c_2^{\dagger} c_3 + \frac{i \sqrt{3}}{2}   \left(c_1^{\dagger} c_3  - c_2^{\dagger} c_4\right) \right] ,
                 \\
                 \hat{H}_{p4} &= \epsilon J \left[ c_2^{\dagger} c_2 + c_3^{\dagger} c_3 + \sqrt{3} \left( c_1^{\dagger} c_3  + c_2^{\dagger} c_4 + i c_1^{\dagger} c_2 - i c_3^{\dagger} c_4  \right) 
                 \right] ,
                 \\
                 \hat{H}_{p5} &=  \epsilon J^2 \left[ \left(i c_3^{\dagger} c_4 - i c_1^{\dagger} c_2 +   \mathrm{h.c.} \right) - \left(c_1^{\dagger} c_3 + c_2^{\dagger} c_4 - \mathrm{h.c.} \right)  \right] .
              \end{split}
              \label{17}
           \end{equation}
           Although the physical meaning of these perturbations is clear in this form, 
           their associated spectral splitting exponent is not directly evident. 
           To apply our theoretical framework, we transform their matrix representations 
           into the basis of the Jordan form using the same similarity transformation. 
           Specifically, we express the perturbation in the original basis 
           as $\hat{H}_{pi} = \psi^{\dag}H_{pi}\psi$ and then apply the transformation 
           to obtain its representation in the Jordan basis: $\epsilon \mathcal{H}_{1}^{pi} = U_4^{-1} H_{pi} U_4$.
           The results are 
           \begin{equation}
              \begin{split}
                 \epsilon \mathcal{H}_1^{p1} &= \epsilon
                   \begin{pmatrix}
                      0 & 0 & 0 & 0 \\
                      0 & 0 & 0 & 0  \\
                      0 & 0 & 0 & 0  \\
                      J^4 & 0 & 0 & 0
                   \end{pmatrix},
                       \\
                 \epsilon \mathcal{H}_1^{p2} &= \epsilon 
                   \begin{pmatrix}
                      0 & 0 & 0 & 0 \\
                      0 & 0 & 0 & 0  \\
                      \frac{1}{2} J^3 & 0 & 0 & 0  \\
                      0 & \frac{1}{2} J^3 & 0 & 0
                   \end{pmatrix},
                  \\
                 \epsilon \mathcal{H}_1^{p3} &= \epsilon 
                   \begin{pmatrix}
                      0 & 0 & 0 & 0 \\
                      0 & 0 & 0 & 0  \\
                      0 & J^2 & 0 & 0  \\
                      0 & 0 & 0 & 0
                   \end{pmatrix},
                   \\
                 \epsilon \mathcal{H}_1^{p4} &= \epsilon  
                   \begin{pmatrix}
                      0 & 0 & 0 & 0 \\
                      0 & J & 0 & 0  \\
                      0 & 0 & J & 0  \\
                      0 & 0 & 0 & 0
                   \end{pmatrix},
                   \\
               \epsilon \mathcal{H}_1^{p5} &= \epsilon
                 \begin{pmatrix}
                      0 & \frac{i}{\sqrt{3}} J & \frac{\sqrt{3}}{6} & 0 \\
                      0 & 0 & 0 & \frac{\sqrt{3}}{6}  \\
                      0 & 0 & 0 & \frac{-i}{\sqrt{3}} J  \\
                      0 & 0 & 0 & 0
                 \end{pmatrix}.
              \end{split}
              \label{23}
           \end{equation}
           According to the framework established in Sec.~\ref{III}, we classify 
           each perturbation and determine its spectral splitting exponent:
           
           (1) $\epsilon\mathcal{H}_1^{p1}$ (Case 2): Its sole nonzero matrix element satisfies $p+q=0$, 
           directly yielding an exponent of $\alpha=1/4$.
           
           (2) $\epsilon\mathcal{H}_1^{p2}$ (Case 4): The two nonzero elements lie on 
           the line $p+q=1$, which immediately gives $\alpha=1/3$.
           
           (3) $\epsilon\mathcal{H}_1^{p3}$ (Case 2): With its only nonzero element at $p+q=2$, 
           the exponent is $\alpha=1/2$.
           
           (4) $\epsilon\mathcal{H}_1^{p4}$ (Case 3): This diagonal perturbation produces a linear splitting, 
           corresponding to $\alpha=1$.
           
           (5) $\epsilon\mathcal{H}_1^{p5}$ (Case 1): Despite having many terms in the original basis, 
           all its nonzero elements are in the upper triangle ($p+q \geq N$), resulting in no splitting of 
           the EP ($\alpha$ is undefined).

\begin{figure}
               \centering
               \includegraphics[scale=0.7]{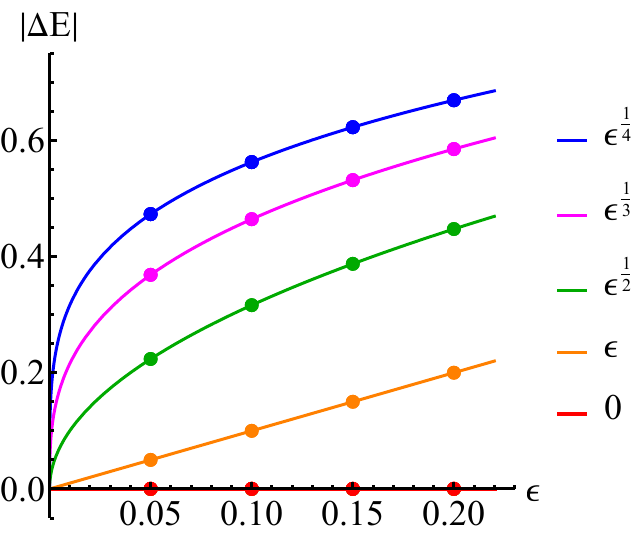}
               \caption{Absolute eigenenergy splitting $|\Delta E|$ for the 4-site non-Hermitian SUSY array 
               under five different perturbations $H_{p1}$ to $H_{p5}$. {\wsx The blue, magenta, green, orange and red dots represent numerical results for $H_{p1}$, $H_{p2}$, $H_{p3}$, $H_{p4}$ and $H_{p5}$, respectively.  Solid curves crossing the dots correspond to the analytical predictions $|\Delta E| = \epsilon^{1/4}$, $\epsilon^{1/3}$, 
               $\epsilon^{1/2}$, $\epsilon$, and $0$, respectively.} Parameters are $\omega_0 = 0$, $\gamma=1$, and $J=1$.}
               \label{fig1}
           \end{figure}

         To demonstrate the predictive power of our theoretical framework, we now calculate 
         the explicit magnitude of the spectral splitting, not just the exponent, using 
         the formulas derived in Sec.~\ref{III}. For the five perturbations, the induced 
         spectral splittings are respectively
           \begin{equation}
              \begin{split}
                |\Delta E|^{(1)} &= \left| \left( \epsilon J^4 \right)^{\frac{1}{4}} \right| = \epsilon^{\frac{1}{4}} \left|J\right|,
                \\
                |\Delta E|^{(2)} &= \left| \left[  \epsilon \left(\frac{1}{2} J^3 + \frac{1}{2} J^3\right) \right]^{\frac{1}{3}} \right| = \epsilon^{\frac{1}{3}} \left|J\right|,
                \\
                |\Delta E|^{(3)} &= \left| \left( \epsilon J^2 \right)^{\frac{1}{2}} \right| = \epsilon^{\frac{1}{2}} \left|J\right|,
                \\
                |\Delta E|^{(4)} &= \epsilon \left|J\right|,
                \\
                |\Delta E|^{(5)} &= 0.
              \end{split}
           \end{equation}
           Setting $J=1$ for simplicity yields the successive results $\epsilon^{1/4}$, $\epsilon^{1/3}$, $\epsilon^{1/2}$, $\epsilon$, and $0$. Fig.~\ref{fig1} compares these analytical predictions (solid curves) with exact numerical results (dots), showing excellent 
           agreement and confirming the accuracy of our theory.

    \section{Discussions and conclusions}\label{V}

           EPs are fundamental spectral singularities in non-Hermitian systems, and 
           understanding their response to perturbations is crucial for developing 
           EP-based applications. At an EP, the Hamiltonian becomes defective, 
           preventing full diagonalization and necessitating a description 
           via its Jordan block structure. In this work, we have developed a general 
           theoretical framework that leverages this Jordan form to analytically 
           predict the spectral splitting exponent $\alpha$ for a perturbed EP. 
           A key finding of our theory is that the exponent $\alpha$ appears to be governed 
           solely by the positions of the nonzero elements in the perturbation matrix. 
           We validated this framework through both analytical and numerical studies of a 
           fourth-order EP under distinct perturbations. The excellent agreement between 
           our analytical predictions and exact numerical results confirms the power and accuracy of our approach.
           Consequently, the theoretical framework developed here provides a powerful design principle 
           for engineering perturbations to achieve a desired spectral response, promising wide 
           applications in non-Hermitian physics.

    \section{Acknowledgments}
          This work is supported by the National Natural Science Foundation of China (Grant No. 12174455) and Guangdong Basic and Applied Basic Research Foundation (Grant No.2023B1515040023)

    \appendix

    \section{A brief review of the Jordan chain structure}
       Consider an $N$th-order EP at energy $E_0$, described by an $N \times N$ 
       non-Hermitian Hamiltonian $H_N$. This Hamiltonian possesses a unique eigenstate $|u_0\rangle$ and, 
       as shown in Ref.~\cite{seyranian2003multiparameter}, $N-1$ associated vectors $\{|u_1\rangle, \dots, |u_{N-1}\rangle\}$ 
       that together form a Jordan chain satisfying:
          \begin{equation}
              \begin{split}
                \left( H_{\rm N} - E_0 \right) &|u_0\rangle = 0,
                \\
                \left( H_{\rm N} - E_0 \right) &|u_1\rangle = |u_0\rangle,
                \\
                &\vdots
                \\
                \left( H_{\rm N} - E_0 \right) &|u_{N-1}\rangle = |u_{N-2}\rangle.
              \end{split}
              \tag{A1}
              \label{A1}
          \end{equation}
       Similarly, the left eigenstate $| v_0 \rangle$ and its $N-1$ left associated vectors $\{|v_1\rangle, |v_2\rangle, \cdots, |v_{N-1}\rangle\}$ together form a Jordan chain satisfying:
           \begin{equation}
              \begin{split}
                \langle v_0|& \left( H_{\rm N} - E_0 \right) = 0,  
                \\
                \langle v_1|& \left( H_{\rm N} - E_0 \right) = \langle v_0| ,
                \\
                & \quad \vdots
                \\
                \langle v_{N-1}|& \left( H_{\rm N} - E_0 \right) = \langle v_{N-2}|.
              \end{split}
              \tag{A2}
              \label{A2}
           \end{equation}
        These two sets of states satisfy  the orthogonal normalization condition:
           \begin{equation}
            \langle v_{i}|u_{j}\rangle = \delta_{i+j,N-1}, 
            \tag{A3}
            \label{A3}
           \end{equation}
        where $i$ and $j$ take values in $\{0,1,\cdots, N-1\}$. 

      \section{Derivation of Eq.~\eqref{10}}
      
      When the matrix $\mathcal{H}_1$ has only one nonzero element $H_1^{p,q}$, the system of equations in Eq.~\eqref{4} simplifies to
            \begin{equation}
               \begin{split}
                  - \lambda s_{N-1} &= 0,
                  \\
                  s_{N-1}  - \lambda s_{N-2} &= 0,
                  \\
                  s_{N-2}  - \lambda s_{N-3} &= 0,
                  \\
                  \vdots    \qquad  \quad &
                  \\
                  s_{N-p+1} - \lambda s_{N-p} &= 0,
                  \\
                  s_{N-p} + \epsilon     s_q H_1^{p,q} - \lambda s_{N-p-1} &= 0,
                  \\
                  s_{N-p-1} - \lambda s_{N-p-2} &= 0
                  \\
                    \vdots    \qquad  \quad &
                  \\
                  s_{2}  - \lambda s_{1} &= 0,
                  \\
                  s_{1}  - \lambda   &= 0.
               \end{split}
               \tag{B1}
               \label{B1}
            \end{equation}
        Here, we have set  $s_0 = 1$. Obviously, the solutions for
        $s_{i}$ are given by 
            \begin{equation}
               s_i =
                  \begin{cases}
                     \lambda^i  &  \mathrm{for} \ 0 \leqslant i \leqslant N-p-1
                     \\
                     0          &  \mathrm{for} \ N-p \leqslant i \leqslant N-1
                  \end{cases}.
                  \tag{B2}
                  \label{B2}
            \end{equation}
            If $p+q < N$ (i.e., $q < N-p$), then $s_q = \lambda^q$. 
            Substituting this into the equation  $s_{N-p} 
            + \epsilon s_q H_1^{p,q} - \lambda s_{N-p-1} = 0$, one immediately obtains
            \begin{equation}
                 \epsilon \lambda^q H_1^{p,q} - \lambda^{N-p}  = 0,
                 \tag{B3}
                 \label{B3}
            \end{equation}
         which is identical to Eq.~\eqref{10} in the main text.

      \section{Derivation of Eq.~\eqref{15}}
      
         In Case 4, when $j < N-1$, Eq.~\eqref{4} can be reduced as
           \begin{equation}
              \begin{split}
               \epsilon s_j H_1^{0,j} - \lambda s_{N-1} &=0,
               \\
               s_{N-1} + \epsilon s_{j-1} H_1^{1,j-1} - \lambda s_{N-2} &=0
               \\
               \vdots \qquad &
               \\
               s_{N-j+1} +  \epsilon s_1 H_1^{j-1,1} - \lambda s_{N-j} &=0
               \\
               s_{N-j} + \epsilon H_1^{j,0} - \lambda s_{N-j-1} &= 0,
               \\
               s_{N-j-1} - \lambda s_{N-j-2} &=0,
               \\
               \vdots \qquad &
               \\
               s_2 - \lambda s_1 &= 0,
               \\
               s_1 - \lambda &= 0.
              \end{split}
              \tag{C1}
              \label{C1}
           \end{equation}
         Similarly, it is easy to find $s_i = \lambda^i$ for 
         $i=\{0,1,\cdots,N-j-1\}$.
         If a more stringent constraint, $j \leqslant \frac{1}{2} \left( N-1 \right)$, is put, then 
         $s_i$ for $i \geqslant N-j$ can be represented as
           \begin{equation}
              \begin{split}
               s_{N-j} =& \lambda^{N-j} + \epsilon H_1^{j,0} ,
               \\
               s_{N-j+1} =& \lambda \left( \lambda^{N-j} + \epsilon H_1^{j,0} + \epsilon H_1^{j-1,1} \right),
               \\
               s_{N-j+2} =& \lambda^2 \left( \lambda^{N-j} + \epsilon H_1^{j,0} + \epsilon H_1^{j-1,1} +\epsilon H_1^{j-2,2} \right),
               \\
               \vdots
               \\
               s_{N-1} =& \lambda^{j-1} \left( \lambda^{N-j} - \epsilon \sum_{k=0}^{j-1} H_1^{j-k,k}  \right).
              \end{split}
              \tag{C2}
              \label{C2}
           \end{equation}
           Substituting the last equation of~\eqref{C2} into the first equation of~\eqref{C1} yields
           \begin{equation}
              \lambda^j \left( \lambda^{N-j} - \epsilon \sum_{k=0}^{j} H_1^{j-k,k}  \right) = 0.
              \tag{C3}
              \label{C3}
           \end{equation}
         The nonzero solutions of this equation are
           \begin{equation}
            \lambda_m = \epsilon^{\frac{1}{N-j}} \left( \sum_{k=0}^{j} H_1^{j-k,k} \right)^{\frac{1}{N-j}} e^{\frac{2 i m \pi}{N-j}},
            \tag{C4}
            \label{C4}
           \end{equation}
         for $m=1,2, \cdots, N-j$, which is the result presented in Eq.~\eqref{15}. 

        {\zbb When $\frac{1}{2}(N-1)<j<N-1$, or equivalently $N-j-1<j<N-1$, there exists an integer $t<j$ 
        such that $s_i \not= \lambda^i$ for $i=j, j-1, \cdots, t$. This means that Eq.~\eqref{C1} cannot be reduced to Eq.~\eqref{C2}. 
        Consequently, it is difficult to solve Eq.~\eqref{C1} analytically. Numerical results in appendix D show that
        a perturbation in this regime typically induces multiple split eigenvalues, all of which scale as $\epsilon^{\frac{1}{N-j}}$. }

      {\wsx
      \section{Perturbations beyond the four types discussed in section $\mathbf{III}$}

        In Section $\mathbf{III}$, we discuss four types of perturbations that can be solved analytically. For more general cases, we conjecture that the spectral splitting exponent is determined by the smallest index $j$ among the nonzero elements of the perturbation matrix, because if we consider all non-zero elements separately, those with the smallest $j$ {\zbb obviously governs} the spectral splitting. 
        \par

        \begin{figure}
          \centering
          \subfigure[]{\includegraphics[scale=0.3]{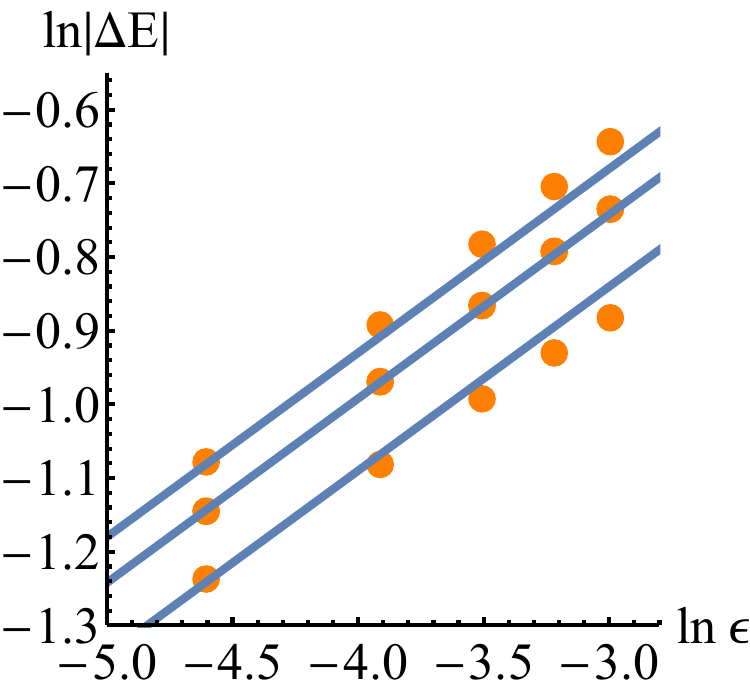}\label{fig2a}}
          \qquad
          \subfigure[]{\includegraphics[scale=0.3]{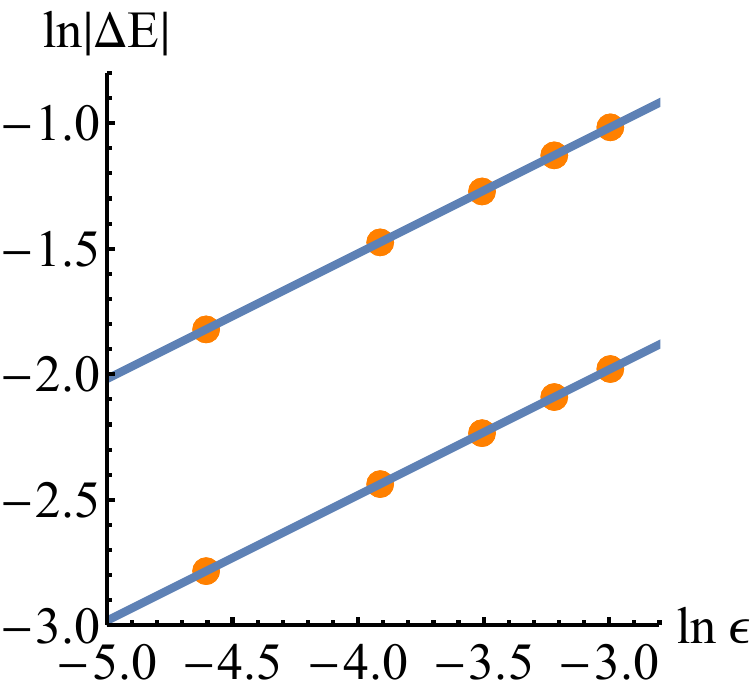}\label{fig2b}}
          \caption{Spectral splitting induced by (a) $\hat{H}_{p6}$ and (b) $\hat{H}_{p7}$. 
          The orange dots represent numerical results. The three lines in (a) 
          correspond to $\ln|\Delta E| = \frac{1}{4} \ln \epsilon + C$ with $C = 0.07$, $0.008$, and $-0.09$, respectively. 
          The two lines in (b) correspond to $\ln|\Delta E| = \frac{1}{2} \ln \epsilon + \frac{1}{2} \ln \frac{3\pm \sqrt{5}}{2}$. 
          Parameters are $\omega_0 = 0$, $\gamma=1$, and $J=1$.}
        \end{figure}
        In this section, we numerically verify this conjecture using the non-Hermitian SUSY array model from Section $\mathbf{IV}$.

         First, we {\zbb investigate the case that the Hamiltonian} $\hat{H}_4$ is perturbed by $\hat{H}_{p6}=\hat{H}_{p1}+\hat{H}_{p2}$. According to Eq.~\eqref{23}, the minimum index $j$ for nonzero elements in {\zbb the} perturbation matrix is $j_{min}=0$ . The numerical results in Fig.~\ref{fig2a} show that the splitting of eigenenergy for all three branches scales as $\epsilon^{\frac{1}{N-j_{min}}}=\epsilon^{\frac{1}{4}}$ for small $\epsilon$, confirming the predicted exponent. 
         
         Next we examine the perturbation
          \begin{eqnarray}
            \hat{H}_{p7} &=& \frac{\epsilon J}{2 \sqrt{3}} \left[ 2\left(c_1^{\dagger} c_2 + c_3^{\dagger} c_4 \right) - i \left(c_1^{\dagger} c_3 - c_2^{\dagger} c_4\right) \right.\nonumber\\
            &&\left.- \sqrt{3} \left( c_1^{\dagger} c_4 - c_2^{\dagger} c_3 \right) \right],
            \label{AppendixD1}
          \end{eqnarray}
         with its corresponding perturbation matrix given by
           \begin{equation}
             \epsilon \mathcal{H}_1^{p7} = {\zbb \epsilon}
               \begin{pmatrix}
                     0 & 0 & 0 & 0 \\
                      J^2 & 0 & 0 & 0  \\
                      0 & J^2 & 0 & 0  \\
                      0 & 0 & J^2 & 0
               \end{pmatrix}.
               \label{AppendixD2}
           \end{equation}
         This case falls outside the four types {\zbb of  perturbations} discussed in Sec.$\mathbf{IV}$, as the index $j$ for all nonzero elements in $\mathcal{H}_1^{p7}$ satisfies $(N-1)/2 < j = N-2 < N-1$. {\zbb Figure \ref{fig2b}} shows numerical results for {\zbb the} spectral splitting induced by $\hat{H}_{p7}$. Both branches scale as $\epsilon^{\frac{1}{N-j}} = \epsilon^{\frac{1}{2}}$, again consistent with our conjecture.
         {\zbb We note that the existence of multiple split eigenvalues with distinct absolute values  explains why the general case is difficult to solve analytically. }

      } 
         

   \bibliography{scale}

\end{document}